\documentclass[twocolumn,pra,aps,showpacs,9pt]{revtex4-1}

\usepackage{amsmath,amssymb,amsfonts,bbm,graphicx}
\usepackage{amstext}

\newcommand{\fref}[1]{Fig.~\ref{#1}}
\newcommand{\sref}[1]{Sec.~\ref{#1}}

\newcommand{\eref}[1]{Eq.~(\ref{#1})}

\newcommand{\Eref}[1]{Equation~(\ref{#1})}

\newcommand{\sandsref}[2]{Secs.~\ref{#1} and \ref{#2}}
\newcommand{\eanderef}[2]{Eqs.~(\ref{#1}) and (\ref{#2})}

\begin{document}

\title{On the classical Schr\"odinger equation}

\author{Albert Benseny}
\affiliation{Quantum Systems Unit, Okinawa Institute of Science and Technology Graduate University, 
Onna-son, Okinawa 904-0495, Japan}
\author{David Tena}
\affiliation{Departament d'Enginyeria Electr\`onica, Universitat Aut\`onoma de Barcelona, 
Bellaterra (Barcelona), E-08193, Spain}
\author{Xavier Oriols}
\affiliation{Departament d'Enginyeria Electr\`onica, Universitat Aut\`onoma de Barcelona, 
Bellaterra (Barcelona), E-08193, Spain}

\begin{abstract}
In this paper, the classical Schr\"odinger equation, which allows the study of classical dynamics in terms of wave functions, is analyzed theoretically and numerically.  First, departing from classical (Newtonian) mechanics, and assuming an additional single-valued condition for the Hamilton\rq{}s principal function, the classical Schr\"odinger equation is obtained. This additional assumption implies inherent non-classical features on the description of the dynamics obtained from the classical Schr\"odinger equation: the trajectories do not cross in the configuration space. Second, departing from Bohmian mechanics and invoking the quantum-to-classical transition, the classical Schr\"odinger equation is obtained in a natural way for the center of mass of a quantum system with a large number of identical particles. This quantum development imposes the condition of dealing with a narrow wave packet, which implicitly avoids the non-classical features mentioned above. We illustrate all the above points with numerical simulations of the classical and quantum Schr\"odinger equations for different systems.
\end{abstract}

\maketitle

\section{Introduction}
\label{sec1}

Since the beginning of quantum theory a century ago, the study of the frontier between classical and quantum mechanics has been a constant topic of debate~\cite{ref3,Zurek03,Giulini96,schlosshauer14}.
In our opinion, the difficulties in the progress of this debate are also due to the deep-rooted use of different languages in classical and quantum mechanics:
while classical system are usually described using trajectories~\cite{classical,goldstein}, quantum descriptions involve wave functions~\cite{landau81}.
These two languages are so different that it is hard (and sometimes misleading) to directly compare trajectories and wave functions.
But despite its pervasiveness, this clash of languages can be avoided by using other readily available formulations, where both theories can be compared on an equal footing.
For instance, Bohmian mechanics provides a deterministic description of quantum systems where particles have definite positions, and their trajectories are \emph{choreographed} by the wave function~\cite{bohm52,Holland93,Oriols12,durr13,ABM_review}.
On the other hand, ensembles of classical particles can be described in terms of waves using the Hamilton--Jacobi formalism, where a classical Schr\"odinger (wave) equation can be derived~\cite{classical,goldstein}.

In this work, we discuss about the classical Schr\"odinger equation. In the first part of the paper, in \sref{sec:HJ}, we will review the path from Newtonian mechanics to the classical Schr\"odinger equation, through the use of the Hamilton--Jacobi formulation~\cite{classical,goldstein}. We will notice that a single-valued condition on the Hamilton\rq{}s principal function is assumed during such development.
This additional assumption implies that the trajectories cannot cross in configuration space, which leads to unexpected non-classical features not present in the Newtonian formalism. As a conclusion, there are inherent non-classical features on the dynamics of the classical Schr\"odinger equation. 

In the second part of the paper, in section \sref{sec:BM}, we analyze different paths to arrive to the classical Schr\"odinger equation from quantum systems by invoking the quantum-to-classical transition within the Bohmian formalism. In particular, we show that the center of mass of quantum systems with a very large number of identical particles tends to behave classically, following the classical Schr\"odinger equation mentioned above.
We notice that this second derivation demands a very narrow wave packet. This additional requirement avoids the inherent non-classical features mentioned in the first part of the paper for the classical Schr\"odinger equation.

In \sref{sec:sim} we provide some numerical results obtained from the quantum and classical Schr\"odinger equations justifying the theoretical discussions mentioned previously. We conclude in \sref{sec:conc}.

\section{From Newtonian mechanics to the classical Schr\"odinger equation}
\label{sec:HJ}

In what follows we present the usual way of arriving to the classical Schr\"odinger equation to describe an ensemble of classical particles starting from Newton's law.
The mathematical derivation for a single-particle system is done in \sref{sec:HJderiva}, while \sref{sec:Sunival} discusses one of the relevant topics of this work: the requirement to the reach the classical Schr\"odinger equation is a single-valued Hamilton's principal function.

\subsection{Hamilton--Jacobi formalism}
\label{sec:HJderiva}

According to classical mechanics~\cite{classical,goldstein}, after specifying the initial position and velocity, the trajectory $X[t]$ of  a particle is fully determined by Newton's second law:
\begin{equation}
\label{newtoncla}
m\frac {d^2 X[t]} {dt^2}  = - \left.\frac {\partial U(x,t)} {\partial x} \right|_{x = X[t]},
\end{equation}
where $m$ is the particle's mass and $U(x,t)$ is a (classical) scalar potential~\cite{footnote1}. It is well known that the previous Newtonian trajectory $X[t]$ is compatible with the Hamilton--Jacobi equation~\cite{classical,goldstein,Oriols12}:
\begin{equation}
\label{hamiltoncla}
H\left(x,\frac{\partial S(x,t;\alpha)}{\partial x},t\right) + \frac{\partial S(x,t;\alpha)}{\partial t} = 0,
\end{equation}
where $H$ is the Hamiltonian of the single-particle system, containing a kinetic energy plus a potential energy:
\begin{equation}
H\left(x,\frac{\partial S(x,t;\alpha)}{\partial x},t\right) = \frac {1} {2m} \left(\frac{\partial S(x,t;\alpha)}{\partial x} \right)^2 + U(x,t).
\end{equation}
The function $S(x,t;\alpha)$ can be identified with Hamilton's principal function (also known as the action~\cite{classical,goldstein}), from which the particle's velocity is derived as
\begin{equation}
v(x,t;\alpha) = \frac {1} {m} \frac{\partial S(x,t;\alpha)}{\partial x} .
\label{eq:v_cl1}
\end{equation}
Here, the additional parameter $\alpha$ is written to emphasize that each particle has its own velocity in the physical point $\{x,t\}$. The velocity in a physical point depends, not only on the point itself, but also on the particle. In other words, two different particles can have two different values of their velocity $v(x,t;\alpha)\neq v(x,t;\alpha')$ (i.e. two different values of Hamilton's principal function $S(x,t;\alpha)\neq S(x,t;\alpha')$) at one unique point of physical space $\{x,t\}$. 
In general, finding the direct solution of \eref{hamiltoncla} is more difficult than solving Newton' law in \eref{newtoncla}. One of the great merits of the Hamilton-Jacobi formalism is its ability to arrive to a wave formulation of classical systems. 
 
Although it describes the dynamics of a classical particle in 1D physical space, the Hamilton's principal function $S(x,t;\alpha_0)$ is defined mathematically as a function in a 2-dimensional (plus time) space. On the contrary, for example, the (wave) function that describes the dynamics of a quantum particle in a 1D physical space, $\psi(x,t)$, is indeed defined in position-time dimensional space. Therefore, in order to connect the classical and quantum world, one is tempted to use the function $S(x,t)$ neglecting its dependence on $\alpha$, to describe classical systems within a wave formulation. We will see that this elimination has dramatic consequences on the ability of the classical Schr\"odinger equation to correctly model classical dynamics. Once we neglect $\alpha$, the particle's velocity in \eref{eq:v_cl1} can be rewritten as:
\begin{equation}
v(x,t) = \frac {1} {m} \frac{\partial S(x,t)}{\partial x} .
\label{eq:v_cl}
\end{equation}  
We emphasize that the new equation \eref{eq:v_cl} implies that the velocity in a point $\{x,t\}$ does not depend on a the particle itself. There is only one possible velocity in $\{x,t\}$ for all particles. Not all ensemble of classical system satisfy this requirement. However, it is a requirement to be able to arrive to a wave description of of a classical ensemble of particles.

Let us assume that we have a classical ensemble of single-particle experiments, each experiment is described by exactly the same Hamiltonian, but with a slightly different initial conditions for the particles. In this case, we can define an ensemble of different initial positions of the particles of different experiments, distributed following $R^2(x,0) \geq 0$.
Each trajectory in this ensemble will evolve according to \eref{newtoncla} so that the function $R^2(x,t)$ describes the distribution of particles, in the phase-space, at any time.
Since all these classical particles will move in a \textit{continuous} way, that is, from one unit of volume of the physical space to another, we can ensure that the ensemble of trajectories accomplishes the following \textit{local} conservation law~\cite{Oriols12}:
\begin{equation}
\label{conservationcla}
\frac{\partial R^2(x,t)}{\partial t} + \frac {\partial } {\partial x} \left(\frac {1} {m} \frac {\partial S(x,t)}{\partial x} R^2(x,t) \right) = 0.
\end{equation}
It can be shown that the two previous (real) equations, Eqs. (\ref{hamiltoncla}) and (\ref{conservationcla}), for $S(x,t)$ and $R(x,t)$ are equivalent to the following (complex) classical Schr\"odinger equation~\cite{Oriols12}:
\begin{align}
\label{1schocla}
i \hbar \frac{ \partial \psi_{cl}(x,t)} {\partial t} &= \left( -\frac {\hbar^2}{2m} \frac{ {\partial}^2 } {\partial x^2} + U(x,t) - Q(x,t) \right) \psi_{cl}(x,t),
\end{align}
where $\psi_{cl}(x,t) = R(x,t) \exp(i S(x,t)/\hbar)$ is defined as a classical (complex) wave function. The additional function $Q(x,t)$ on \eref{1schocla} is the so-called quantum potential:
\begin{equation}
\label{quantumpotential}
Q(x,t)
= - \frac {\hbar^{2}} {2 m} \frac{1}{R(x,t)}\frac { {\partial}^2 R(x,t)}{\partial x^2},
\end{equation}
The reader can be, perhaps, surprised that this function, appearing in a purely classical context, is referred to as a quantum entity.
The justification is based on historical arguments~\cite{Oriols12,bohm52}, as it appeared first in the hydrodynamic or Bohmian interpretation of quantum mechanics.

\subsection{On the single-valued $S(x,t)$}
\label{sec:Sunival}

In the transition from Newton's equation, \eref{newtoncla}, to the classical Schr\"odinger equation, \eref{1schocla}, we have made an additional (and usually unnoticed) but critical assumption.
We have assumed that Hamilton's principal function $S(x,t;\alpha)$ becomes a single-valued function (having all velocities of an ensemble of particles the same velocity in each position $\{x,t\}$). 

Let us explain with detail the critical problem mentioned here.
Newton's second law can be used to describe any trajectory occurring in a given experiment. We can therefore use it to describe one experiment defined by a trajectory $X[t]$ and another realization of the same experiment with a different trajectory $X\rq{}[t]$.
However, a problem arises when we define $\psi_{cl}(x,t) = R(x,t) \exp(i S(x,t)/\hbar)$, as we force the trajectories' velocities of $X[t]$ and $X\rq{}[t]$ to fulfil \eref{eq:v_cl} for the same action function.
The use of the wave function implicitly assumes that $S(x,t)$ is single-valued, which in turn implies that different trajectories (described by the same $S(x,t)$) will not cross in the physical (configuration) space. However, in Newtonian mechanics, trajectories actually cross with each other.

This fact is intimately related with how a distribution of positions and momenta in a classical ensemble is represented with the classical Schr\"odinger equation.
Not any general probability distribution of positions and momenta, e.g., $R(x,p,t)^2$, can be represented with this formalism, as the it imposes a very strong restriction on the momenta.
In order for the classical wave function to have a well-defined single-valued phase, 
each position is only allowed to have a single momentum (the momentum is obtained from the derivative of the phase of the wave function).
Therefore, only those distributions where all the particles in a particular position have the same velocity can be described by the classical Schr\"odinger equation, i.e., $R(x,p,t)^2 = R(x,p(x),t)^2 = R(x,t)^2$.
However, the formalism is not exempt of problems even in the case where these particular distributions are used, as we explain in the following and illustrate in the simulations in \sref{sec:sim}.

\begin{figure}
\centering
\centerline{\includegraphics[width=0.95\columnwidth]{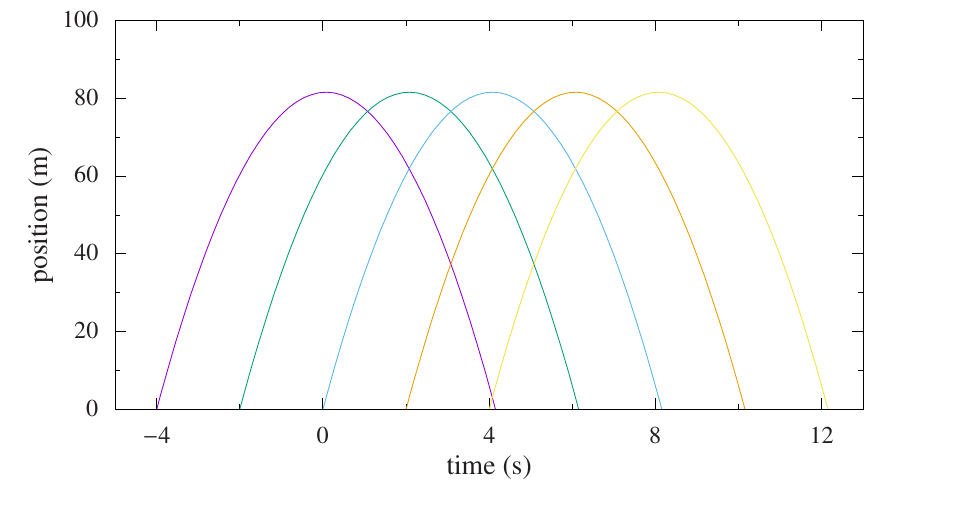}}
\caption{
Trajectories of a classical particle given by \eref{eq:freefall} for different initial positions $X[0]$ with $v_0 = 40$ m/s and $g= 9.81$ m/s$^2$.
}
\label{fig:david}
\end{figure}

To understand better the physical implications of this additional requirement of single-valuedness of $S(x,t)$, let us illustrate the problem in very simple system where this issue becomes relevant: a particle falling under the action of a constant potential.
Consider a particle in a one-dimensional physical system, under a potential $V(x) = m g x$, where $m$ is the mass of the particle and $g$ is a positive constant, e.g. gravity.
From Newton's law we get that the acceleration is constant and equal to $-g$ such that the trajectory for a particle with initial position $X[0]$ and velocity $v_0$ is
\begin{equation}
X[t]=X[0]+v_0 t - \frac{1}{2} g t^2 .
\label{eq:freefall}
\end{equation}
Some trajectories starting at different initial positions $X[0]$ are shown in \fref{fig:david}
The relevant point to our discussion is the presence of trajectories with different velocities (one positive and one negative) at a unique point in configuration space (see, for example,  \{$x=60$ m, $t=4$ s\} in \fref{fig:david}). This duality of the velocity in a single point is not possible when we assume a single-valued function $S(x,t)$.

Let us now use the Hamilton--Jacobi formalism~\cite{classical,goldstein} to tackle the same simple problem.
Since the Hamiltonian of the problem is $H(x)=\frac{p^2}{2m} + mgx$
and does not explicitly depend on time, then $S(x,t;\alpha)= W(x) - E_{\alpha} t$, and the Hamilton--Jacobi equation can be partially solved giving for the momentum of the particle~\cite{footnote2}.
\begin{equation}
\frac{\partial W(x;\alpha)}{\partial x} = \frac{\partial S(x,t;\alpha)}{\partial x} = \pm \sqrt{(E_{\alpha}-mgx)2m}.
\label{eq:3}
\end{equation}
where $E_{\alpha}$ depends on the initial conditions of the particle, and in this case it corresponds to the total mechanical energy of the system.
The two possible signs in \eref{eq:3} imply that (at a particular point of $\{x,t\}$) one particle $\alpha$ can have a positive velocity while going up, and another particle $\alpha'$ can have a negative velocity when going down.
Therefore, the Hamilton--Jacobi formulation is capable of describing the trajectories in \fref{fig:david} with a multivalued solution for $S(x,t;\alpha)$ without any problem. 

Now suppose we want to describe an ensemble of trajectories with a classical wave function $\psi_{cl}$.
At each position, their distribution will be given by some single-valued $R(x,t)$, and their velocity described by some single-valued $S(x,t)$.
As we force $S(x,t)$ to be single-valued in \eref{1schocla}, the trajectories obtained from the classical Schr\"odinger equation will deviate from the (true) ones given by Newton's law in order to avoid crossing.
We will see many examples of this in \sref{sec:sim}. 
The conclusion of this discussion are quite dramatic. In the construction of the classical Schr\"odinger equation, we use $S(x,t)$ as a single-valued function, neglecting the parameter  $\alpha$ that provide a velocity for each particle at each physical point. This additional requirement on $S(x,t)$ produces inherent non-classical features on the classical Schr\"odinger equation, not present in the Newtonian formulation of the problem.

As a final note, we must add that the single-valued $S(x,t)$ implies non-crossing property for the trajectories. One can understand the non-crossing requirement as if the different \lq\lq{}classical\rq\rq{} trajectories in different experiments~\cite{footnote3} have some type of \lq\lq{}non-crossing interaction\rq\rq{} among them. An \lq\lq{}non-crossing interaction\rq\rq{} between the trajectories of different experiments is something unexpected classically, but it is actually the essential characteristic of quantum mechanics, the so-called ``quantum wholeness'' where the dynamics of one particle in one experiment is governed by the probability distribution  $R^2(x,t)$ of the ensemble of positions in all experiments~\cite{wholeness}.

\section{From Bohmian mechanics to the classical Schr\"odinger equation}
\label{sec:BM}

This section contains a completely different way of arriving to the classical Schr\"odinger equation. We depart from the quantum world by invoking the quantum-to-classical limit in Bohmian mechanics.
We start by introducing the basic ideas of Bohmian mechanics in \sref{sec:BMintro}, followed by a discussion of classicality in single-particle Bohmian mechanics in \sref{sec:BMclass} and its drawbacks.
In \sref{sec:BMcom}, we briefly explain a natural path to obtain classical behavior for the center of mass of a quantum system with a very large number of identical particles, and in \sref{sec:BMnarrow} we discuss that this path imposes a condition on the shape of the quantum wave packet to reach classical dynamics.

\subsection{Brief introduction to Bohmian mechanics}
\label{sec:BMintro}

We present here a single-particle description of a quantum (sub)system according to the Bohmian formalisms. More detailed derivations, also including discussions on many-particle systems can be found elsewhere~\cite{bohm52,Holland93,Oriols12,durr13,ABM_review}. Our starting point is the single-particle Schr\"{o}dinger equation:
\begin{align}
i \hbar \frac{\partial \psi(x,t)}{\partial t} = - \frac{\hbar^2}{2m} \frac{ \partial^2 \psi(x,t)}{\partial x^2} + U(x,t) \psi (x,t) . 
\label{1scho}
\end{align}
By casting into it the polar form of the wave function,
\begin{align}
\psi(x,t)=R(x,t)e^{i S(x,t) / \hbar}
\end{align}
one arrives at two equations. The first equation is:
\begin{align}
0 &= \frac{\partial S(x,t)}{\partial t} + \frac{1}{2m} \left( \frac{\partial S(x,t)}{\partial x} \right)^2 + U(x,t) + Q(x,t) , 
\label{hamilton} 
\end{align}
This \Eref{hamilton} is the so-called quantum Hamilton--Jacobi equation because of its similarity with the (classical) Hamilton--Jacobi equation in \eref{hamiltoncla} but with one additional term, the quantum potential defined in \eref{quantumpotential}, which accounts for the quantum features of the system~\cite{bohm52,Holland93,Oriols12,durr13,ABM_review}. From the similarities between the classical \eref{hamilton} and the quantum \eref{hamiltoncla}, a velocity \emph{guiding} field can be defined as~\cite{footnote4}:
\begin{align}
v(x,t) &= \frac{1}{m} \frac{\partial S(x,t)}{\partial x} ,
\label{bohmvel} 
\end{align}
Once the velocity is defined, a quantum trajectory (initially located at $X[0]$), can be defined as easily as in classical mechanics as:
\begin{align}
X[t] &=X[0]+ \int_0^t v(X[t^\prime],t^\prime) dt^\prime .
\label{trajint}
\end{align}
The second equation obtained from \eref{1scho} is:
\begin{align}
0 &= \frac{\partial R(x,t)^2}{\partial t} + \frac{1}{m} \frac{\partial}{\partial x} \left( R(x,t)^2 \frac{\partial S(x,t)}{\partial x} \right) 
\label{conservation}
\end{align}
This \eref{conservation} is a continuity equation for the conservation of a probability density $R^2(x,t)$. It ensures \emph{equivariance} for an ensemble of trajectories: if the initial positions of such ensemble are distributed according to $R^2(x,0) = |\psi(x,0)|^2$,
the positions at any other future time $t$ will be distributed according to $R^2(x, t)= |\psi(x,t)|^2$. 

The selection of the initial distribution of particles according to $R^2(x,0) = |\psi(x,0)|^2$ is based on a \emph{quantum equilibrium hypothesis}. Some authors argue that it is an additional postulate of the theory, while others argue that this is just a particular fact verified in virtually all systems under study~\cite{bohm52,Holland93,Oriols12,ABM_review}

In summary, the development performed up to here in \sandsref{sec:HJ}{sec:BM} provides a common language for classical and quantum theories, in terms of either wave functions or trajectories. We emphasize that one has to compare either classical and quantum wave functions or classical and quantum ensembles of trajectories (not a single classical trajectory with a quantum wave function).  The presence of $Q(x,t)$ in the quantum versions of the Hamilton--Jacobi in  \eref{hamilton}, implies that each Bohmian trajectory of a single-particle experiment depends explicitly on the shape of $R(x,t)$.
On the contrary, each classical trajectory can be computed (from Newton's law or the Hamilton--Jacobi equation) independently of the shape of the classical ensemble. Therefore, the differences between quantum and classical ensembles of trajectories do not correspond to differences between waves and particles, because both descriptions can be used to both kinds of systems. The difference is on the dependence of each trajectory on $R(x,t)$ or not. We note, finally, that as seen in \sref{sec:Sunival}, the classical Schr\"{o}dinger equation forces somehow a dependence of the trajectories on $R(x,t)$ because \eref{1schocla} imposes the additional non-crossing condition for the ensemble of trajectories.

\subsection{Classicality in single-particle Bohmian mechanics}
\label{sec:BMclass}

It can be easily demonstrated using \eref{conservation} and \eref{hamilton} that a Newton-like equation can be developed for the quantum (Bohmian) trajectories~\cite{bohm52,Oriols12}:
 \begin{equation}
\label{newton}
m\frac {d^2 X[t]} {dt^2} = \left. -\frac {\partial} {\partial x} \Big[U(x,t) + Q(x,t) \Big] \right|_{x = X[t]}.
\end{equation}
Here, again, we confirm that each unique quantum (Bohmian) trajectory depends on the quantum potential $Q(x,t)$ that in turns depends on $R(x,t)$. 

In 1964, Nathan Rosen~\cite{Rosen64} used the Bohmian formalism to develop the conditions needed to reach a classical regime for a one-particle system. We follow here his single-particle arguments. 
Comparing \eanderef{newtoncla}{newton} it is easy to see that Bohmian trajectories with classical behavior can be obtained by imposing~\cite{Rosen64,sevensteps,Allori09}
\begin{equation}
\label{quantumpotentialzero}
\frac{\partial Q(x,t)}{\partial x} = 0 ,
\end{equation}
i.e., that $Q(x,t)$ is constant everywhere.
This strategy to reach a classical equation, however, can be quite problematic because most of the time it is incompatible with a well-defined wave function solution of \eref{1scho}.
The reason is because we have three equations, Eqs. (\ref{hamilton}), (\ref{conservation}) and (\ref{quantumpotentialzero}), imposed on only two unknowns, $R(x,t)$ and $S(x,t)$.
The number of unknowns is sometimes artificially increased in the literature by also assuming the (classical) potential $U(x,t)$ as an unknown to be fixed by these equations.
Some very exotic solutions can be found where quantum and classical solutions are the same~\cite{Mako02}.
A simple example is a plane wave of momentum $\hbar k$ in free space, $U(x,t)=0$, such that $R(x,t)=1$ and $S(x,t)=k x$.

Another, more phenomenological, way of arriving to the classical Schr\"odinger equation was presented by Richardson \emph{et al.}~\cite{richardson_nonlinear_2014}.
They define $\epsilon$, a \emph{degree of quantumness}, into the Schr\"odinger equation such that it reads
\begin{align}
i \hbar \frac{\partial \psi(x,t)}{\partial t} = \left[- \frac{\hbar^2}{2 m} \frac{\partial ^2}{\partial x^2} + V(x,t) - (1 - \epsilon) Q \right] \psi(x,t) .
\end{align}
For $\epsilon = 1$ this equation becomes the Schr\"{o}dinger equation and for $\epsilon = 0$ it becomes the classical Schr\"{o}dinger equation.
This equation has the virtue that for intermediate values of $\epsilon$ allows the study of the appearance of quantum behavior as $Q$ becomes more and more relevant (and thus the system becomes more classical). However, an explanation of the origin or physical meaning of $\epsilon$ is still missing in the literature.

\subsection{Bohmian mechanics for the center of mass}
\label{sec:BMcom}

In what follows we summarize our recent work on the quantum-to-classical transition through a generalization of Rosen's attempt~\cite{Rosen64} by analyzing the dynamics of the center of mass of a many-particle system with a large number of identical particles~\cite{ref3}.
We show that classicality appears as a \emph{natural} quantum limit of the center of mass of most macroscopic objects, i.e. the effect of the quantum potential associated to the center of mass becomes negligible while still retaining a well-defined wave function.

We start by providing some simple arguments on why it is better to formulate the quantum-to-classical transition in terms of the center of mass rather than in terms of individual particles. First, it is important to notice that a center of mass of a quantum object can have a classical behavior, while the fundamental particles that form the object (and used to compute the center of mass) are still fully quantum. Second, It can be shown that the quantum potential of the center of mass is the sum of the quantum potentials of the particles of the quantum object (which can take positive and negative values). Therefore, one can expect that the value of the quantum potential averaged over a large number of particles will tend to become small (even if the effect of the quantum potential in each particle is not negligible). Let us emphasize the quantum potential acting on a particle is, in general, much more complex than the classical potential. Therefore, the averaged classical potential over the object will basically remain equal to the classical potential assigned to one particle, while the averaged quantum potential tends to be much smaller. Finally, by looking the quantum-to-classical transition through the center of mass of  a large number of particles, we are imposing a "natural" type of coarse-grained procedure: only macroscopic objects (whose center of mass involves a very large number of particles) will become classical.

After these qualitative arguments, next we briefly summarize our work of ref.~\cite{ref3}. We consider a quantum system defined by the many particle wave function $\Psi(x_1,....,x_N)$ where $x_i$ for $i=1,\ldots,N$ are the positions of $N$ particles with mass $m$. We consider a change of variables, from $\{x_1,x_2,..,x_N\}$ to the center of mass $x_{cm}$ and a set of relative coordinates $\vec{y}=\{y_2,..,y_N\}$,
\begin{align}
\label{eq:si}
x_{cm} &= \frac{1}{N} \sum_{i=1}^{N}x_i, 
\qquad
y_j = x_j - \frac{(\sqrt{N}x_{cm}+x_1)} {\sqrt{N} + 1} ,
\end{align}
By rewriting the many-particle Schr\"odinger equation as a function of the center of mass $x_{cm}$ and a set of relative coordinates $\vec{y}=\{y_2,..,y_N\}$, one arrives at the equation for $\Psi(x_{cm},\vec y,t)$~\cite{ref3}
\begin{align}
\label{eq:condi}
i \hbar \frac{\partial \Psi}{\partial t} = \left( - \frac{\hbar^2}{2 M}\frac{\partial^2}{\partial x_{cm}^2}  -\frac{\hbar^2}{2m}  \sum_{j = 2}^N  \frac{\partial^2}{\partial y_j^2} + U(x_{cm},\vec{y},t) \right) \Psi ,
\end{align}
with $M=N m$. Following similar development as the ones above, the force acting on the (Bohmian) center of mass trajectory is given by
\begin{align}
\label{realcond}
M \frac{d^2 X_{cm}[t]}{dt^2} &= -\left. \frac{\partial }{\partial x_{cm}} \left(U(x_{cm},\vec{y},t) + Q_{cm}  + \sum_{j=2}^N  Q_{j} \right) \right|_{x_{cm}=X_{cm}[t] \atop \vec y=\vec Y[t]}
\end{align}
with the quantum potentials
\begin{align}
Q_{cm} \equiv -\frac{\hbar^2}{2M} \frac{1}{R} \frac{\partial^2 R}{\partial x_{cm}^2 } ,
\qquad
Q_{j} \equiv -\frac{\hbar^2}{2m} \frac{1}{R} \frac{\partial^2 R}{\partial y_j^2 } .
\end{align}

The wave function $\Psi(x_{cm},\vec y,t)$ depends on the center of mass $x_{cm}$ and it includes correlations with all relative coordinates $\vec y$.
A single particle wave function of the center of mass can be easily constructed from Bohmian mechanics through the concept of conditional wave function~\cite{Oriols12,durr13,ABM_review}.
The conditional wave function for the center of mass can be obtained by evaluating the many-particle wave function, $\Psi(x_{cm},\vec y,t)$, at the positions of the Bohmian trajectories for the rest of degrees of freedom,
e.g., $\psi(x_{cm},t) \equiv \Psi(x_{cm},\vec{Y}[t],t)$.
Then, by construction, the conditional wave function will yield the same Bohmian velocities for the center of mass as would the full wave function, and by extension the same trajectory.
Following Refs.~\cite{ref3,Oriols07}, the equation of motion for the conditional wave function is
 \begin{align}
\label{eq:conditional}
i \hbar \frac{\partial \psi}{\partial t} = &- \frac{\hbar^2}{2 M}\frac{\partial^2 \psi}{\partial x_{cm}^2}  -\frac{\hbar^2}{2m}  \sum_{j = 2}^N  \left.\frac{\partial^2  \Psi(x_{cm},\vec{y},t)}{\partial y_j^2} \right|_{\vec{y}=\vec{Y}[t]}\nonumber\\ &- i\hbar \sum_{j = 2}^N  v_j^h[t] \left. \frac{\partial \Psi(x_{cm},\vec{y},t)}{\partial y_j}\right|_{\vec{y}=\vec{Y}[t]}+ U(x_{cm},\vec{y},t)\psi ,
\end{align}
a particular case single-particle (conditional) wave equation of the type
\begin{align}
\label{conditional}
i\hbar\frac{\partial \psi}{\partial t} &= \left[-\frac{\hbar^2}{2m}\frac{ \partial}{\partial x^2}+U(x,\vec Y[t],t) \right.
\nonumber \\ &\qquad 
\left. \vphantom{\frac{ \partial}{\partial x^2}} +G_{a}(x_a,\vec Y[t],t)+ i J_{a}(x_a,\vec Y[t],t) \right] \psi .
\end{align}
The additional potentials $G_{a}(x_a,\vec Y[t],t)$ and $J_{a}(x_a,\vec Y[t],t)$ are defined through the many-body wave function~\cite{Oriols07}.
The relevant point here in this many-particle generalization of Rosen\rq{}s attempt (see \sref{sec:BMclass}) is that the new equation \eref{eq:conditional} has the additional environment degrees of freedom $G_{a}(x_a,\vec Y[t],t)$ and $J_{a}(x_a,\vec Y[t],t)$ included.
Now, contrarily to the single-particle case, the number of equations and unknowns in the quantum-to-classical transition is well balanced and it is possible to recover classical dynamics.

To obtain a classical behavior for the (quantum) center of mass (not other degrees of freedom) using \eref{realcond}, an additional requirement is needed, 
\begin{equation}
\frac {\partial U(x_{cm},\vec{y},t)} {\partial x_{cm}}  \gg \frac {\partial Q_{cm}(x_{cm},\vec{y},t)} {\partial x_{cm}}  + \sum_{j=2}^N  \frac {\partial Q_{j}(x_{cm},\vec{y},t)} {\partial x_{cm}},
\label{condition6}
\end{equation}
along the path $X_{cm}[t]$. Therefore, we have again three real equations  to get a classical solution and, now, we have three unknowns $R(x_{cm},t)$, $S(x_{cm},t)$ and $G(x_{cm},t)$~\cite{footnote5}. Then, when \eref{condition6} is satisfied, we arrive to the classical Schr\"odinger equation in \eref{1schocla} for the center of mass of a macroscopic system $x_{cm}$.

In summary, it is possible to obtain classical dynamics for the center of mass of a (very large) quantum system in a single experiment~\cite{ref3}.
This result can also be interpreted as a single-experiment generalization of the well-known multiple-experiment Ehrenfest theorem.

\subsection{Implications of a narrow $R(x_{cm},t)$}
\label{sec:BMnarrow}

In our recent work~\cite{ref3}, we presented the conditions required for a quantum system of identical particles to ensure that its center of mass $x_{cm}$ has a classical behavior. The qualitative arguments are mentioned in \sref{sec:BMcom} and the mathematical development can be found in the appendix of our Ref. ~\cite{ref3}. 
In particular, we discussed when and why it is reasonable to expect that the spatial derivatives of the classical potential in \eref{condition6} are expected to be larger than those of the quantum potentials.
We notice that a relevant requirements for satisfying the condition in \eref{condition6} is that the (center of mass conditional) wave function has to be very narrow. We argue that the requirement of a narrow wave packet is a natural requirement for the center of mass of a quantum object with a large number of particles.

By using the double nature of $\Psi$, as a probability distribution and as a guiding field~\cite{Holland93,Oriols12,durr13},
one can roughly anticipate the shape of such (center of mass conditional) wave packet. 
We consider that we repeat the experiments with the \lq\lq{}same\rq\rq{}~\cite{footnote6} macroscopic object. According to the quantum equilibrium hypothesis mentioned in \sref{sec:BMintro}, each of these experiments will have a different distribution of Bohmian particles. However, all these distribution of Bohmian particles will imply very similar center of mass. We can reasonable assume that each position $X_j$ has a distribution with a standard deviation $\sigma_j$. When $N \rightarrow \infty$, (a kind of central limit theorem analyzed in the appendix of our Ref. ~\cite{ref3})  the distribution $\rho$ of the center of mass can be roughly approximated by a  normal distribution (around a central value $\bar{x}$),
\begin{align}
R(x_{cm}) \approx \frac{1} {\sqrt{2 \pi} \sigma_{cm}} \exp\left(-\frac{(\bar x-x_{cm})^2}{2\sigma_{cm}^2}\right).
\label{gaus}
\end{align}
and the dispersion of the center of mass, $\sigma_{cm}$, follows~\cite{ref3} then
\begin{align}
\sigma_{cm}^2 = \frac {1} {N^2} \sum_{i=1}^{N}\sigma_i^2 \equiv \frac {\sigma^2} {N}
\label{variance}
\end{align}
We clearly see from \eref{variance} that when the number of particles of the macroscopic object increases, the wave packet becomes narrower. 

The combination of the classical  Schr\"odinger equation and a narrow wave packet has some unexpected beneficial effects in our final goal of describing the quantum-to-classical transition. First, as we will see in \sref{sec:sim}, a narrow wave packet governed by the classical  Schr\"odinger equation tends to remain narrow all the time (this is due to the non-linearity of this equation). This effect in addition implies that there is a negligible (quantum) uncertainty on the values of the position and momentum of the center of mass.

In \sref{sec:sim} we will see that these narrow shape of the wave packet
imply, in fact, that the solutions of the classical Schr\"odinger equation in \eref{1schocla} are fully classical, where the problems due to the single-valued $S(x,t)$ becomes irrelevant.

\section{Numerical simulations for the quantum and classical Schr\"odinger equations}
\label{sec:sim}

In this section, we will compare three types of trajectories associated to three equations of motions discussed in this work for the same physical problem (i.e. the same Hamiltonian). First, the quantum/Bohmian trajectories obtained from the (quantum) Schr\"odinger equation. Second, the trajectories obtained from the classical Schr\"odinger equation (CSE). Third, the (actual) classical trajectories, i.e., obtained from Newtonian mechanics.
We will check several physical problems with several potentials $V$ and initial wave functions, and we will use units where $\hbar = m = 1$ throughout.

The evolution of the wave functions is performed by means of a spectral split-step integration~\cite{bauke_accelerating_2011} of the (classical and quantum) Schr\"odinger equations.
However, some care needs to be taken into account when integrating the CSE as the quantum potential can easily be the source of numerical noise.
Because of its $R^{-1}$ dependence, the numerically-calculated $Q$ is very prone to having (unphysical) quick oscillations and discontinuities where $R$ is small.
Therefore, before calculating $Q$ using finite differences for the second derivative, $R$ is smoothed out using a $n$-point smoothing function (depending on the case).

We consider an ensemble of (classical or quantum) $N_E$ experiments. The trajectories in each experiment are selected according to the modulus squared of the initial wave function $|\psi(x,t=0)|^2$. In order to obtain nicely spaced trajectories, we fix the initial position set $X_k(0)$ according to 
\begin{align}
\int_{X_k(0)}^{X_{k+1}(0)} |\psi(x,t=0)|^2 dx = \frac{1}{N_E}
\end{align}
for $k=1,\ldots,N_E-1$. The evolution of each quantum or CSE trajectory is obtained by integrating \eref{trajint} with the velocity in \eref{bohmvel} (or \eref{eq:v_cl}).

\subsection{Single wave packet in free space}

\begin{figure}
\centerline{\includegraphics[width=0.99\columnwidth]{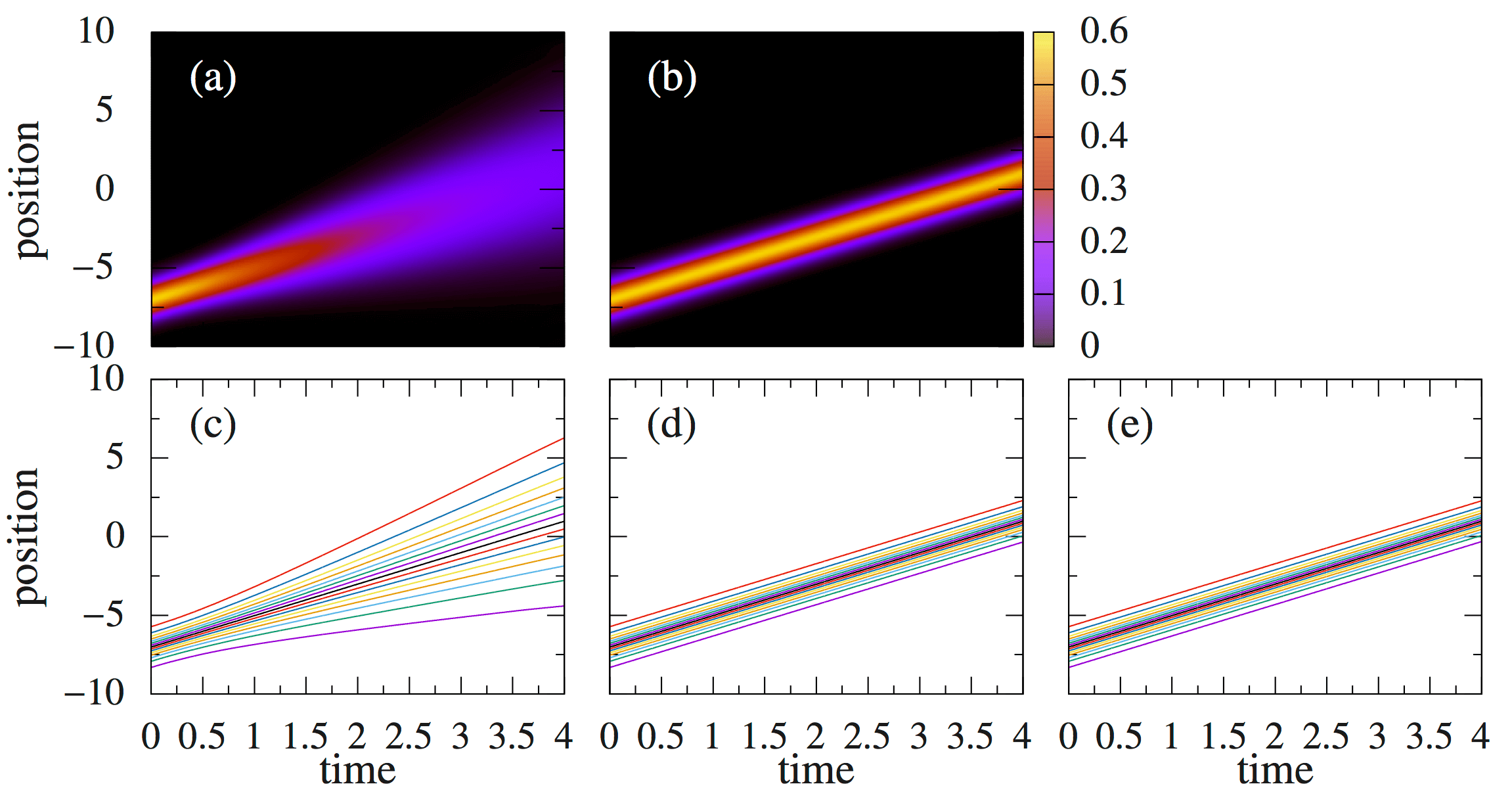}}
\caption{Simulations for a wave packet in free space ($V=0$) with an initial momentum.
The initial wave function is given by \eref{eq:gauss} with $\sigma = 1$, $x_0 = -7$, and $k_0 = 2$.
Top row shows $|\psi(x,t)|^2$ for the integration of the (a) quantum and (b) classical Schr\"odinger equations.
Bottom row shows the trajectories corresponding to the (c) quantum, (d) CSE and (3) classical dynamics.
\label{fig:fs_gauss}
}
\end{figure}

The first case one can consider is free-space dynamics, i.e. $V=0$.
For simplicity we will consider an initial Gaussian wave packet of width $\sigma$, centered around $x_0$ and with an initial momentum $k_0$.
\begin{align}
\Psi_G(x) = \frac{1}{\sqrt{\sigma\sqrt{\pi}}} e^{ -\frac{(x-x_0)^2}{2 \sigma^2} } e^{i k_0 x} .
\label{eq:gauss}
\end{align}

The results of the simulations for $\sigma = 1$, $x_0 = -7$, and $k_0 = 2$ are shown in \fref{fig:fs_gauss}.
Let us start by comparing the wave dynamics.
Both the classical and quantum wave packets propagate forward with a mean velocity $k_0$.
However, while the quantum wave packet expands as it propagates forward, \fref{fig:fs_gauss}(a), the CSE one retains its shape during the whole simulation, \fref{fig:fs_gauss}(b).
Because of this, the quantum trajectories do not follow straight lines and separate during the evolution, \fref{fig:fs_gauss}(c), while the CSE ones follow straight lines, \fref{fig:fs_gauss}(d), as one would expect for a free particle with fixed momentum, as shown in \fref{fig:fs_gauss}(e).
This separation of the quantum trajectories (and the non-expansion of the classical wave packet) can be seen due to the presence (or absence) of the quantum potential (in this case, an inverted parabola centered on the wave packet) in their equations of motion.
In this particular case, since the different trajectories in the classical case do ``interact'' among them (i.e. they do not tend to cross) the problem of having a single-valued $S(x,t)$ does not appear here; and the CSE trajectories perfectly match the classical ones.

\subsection{Wave packet interference}

\begin{figure}
\centerline{\includegraphics[width=0.99\columnwidth]{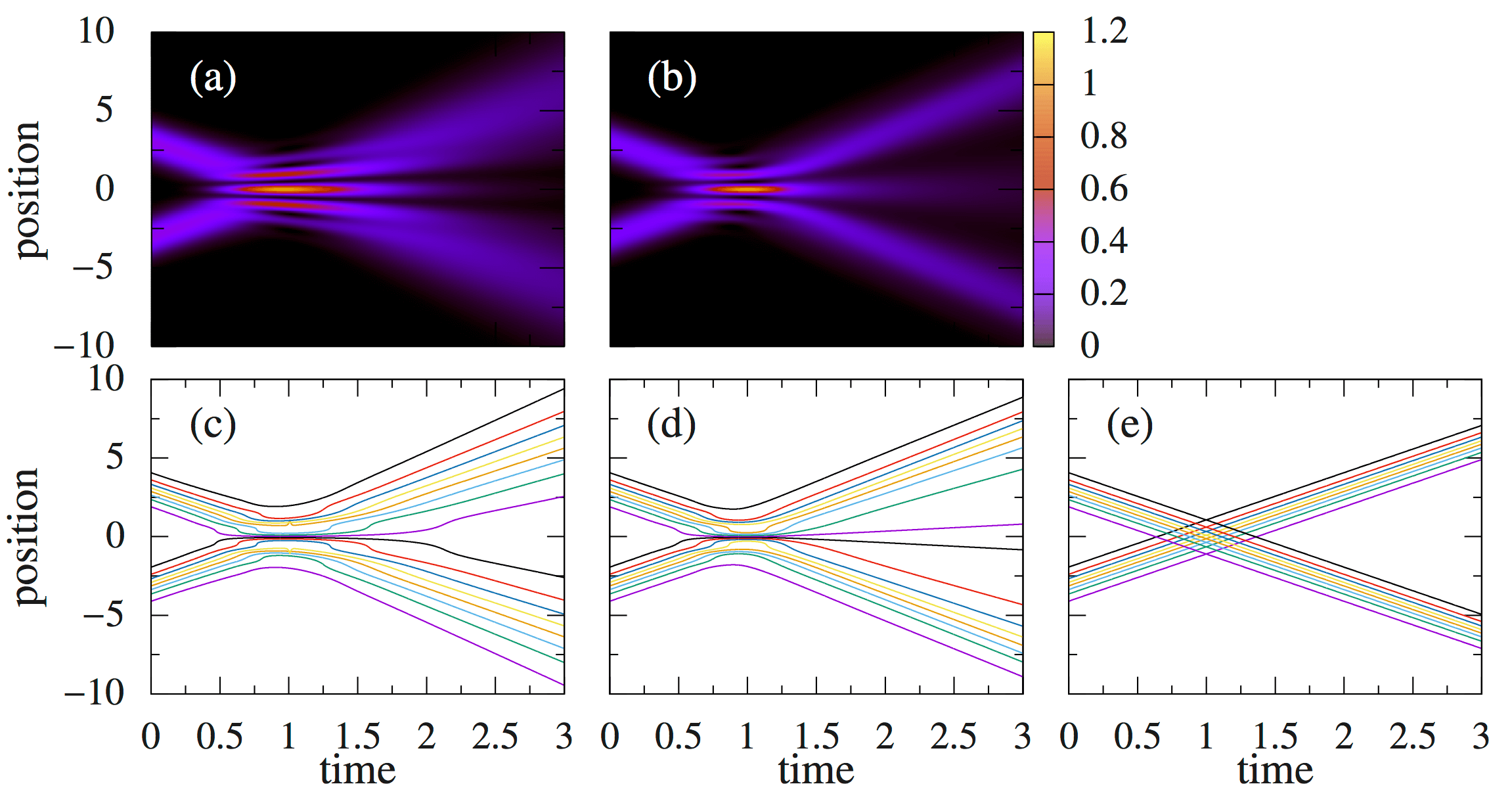}}
\caption{
Same as \fref{fig:fs_gauss} but for two wave packet colliding in free space ($V=0$).
The initial wave function is given by \eref{eq:twogauss} with $\sigma = 1$, $x_0 = 3$, and $k_0 = 3$.
\label{fig:fs_twogauss}
}
\end{figure}

A more complicated case where one can expect trajectories to cross is considering the collision of two wave packets, i.e., an initial wave function
\begin{align}
\Psi_I(x) = \frac{1}{\sqrt{\sigma\sqrt{\pi}}} e^{ -\frac{(x+x_0)^2}{2 \sigma^2} } e^{i k_0 x} + \frac{1}{\sqrt{\sigma\sqrt{\pi}}} e^{ -\frac{(x-x_0)^2}{2 \sigma^2} } e^{- i k_0 x} .
\label{eq:twogauss}
\end{align}
i.e., the superposition of two Gaussian wave packets centered at $\mp x_0$ with momenta $\pm k_0$. The results are shown in \fref{fig:fs_twogauss}.
This is a similar case to the one studied in Ref.~\cite{richardson_nonlinear_2014}, but because its authors did not consider initial momenta, they could not see the interference and shape of the wave packets after them.

For both the quantum and classical wave functions, \fref{fig:fs_twogauss}(a,b), before the two wave packets interact (for $t\lesssim 0.5$), we can see each one doing similar dynamics to the previous case.
However, when they find each other we can see interferences happening which disappear once the packets fly past each other.
After the interferences the wave packets continue with the same dynamics as before the collision.
Since each trajectory represents an independent single-particle experiment, Newtonian mechanics will yield straight (crossing) trajectories, \fref{fig:fs_twogauss}(e).
However, because of the single-valuedness of $S(x,t)$ discussed in \sref{sec:Sunival}, the CSE trajectories, \fref{fig:fs_twogauss}(d), do not follow this straight paths and bounce off each other as in the quantum case, \fref{fig:fs_twogauss}(c).
This effects can also be seen in the classical wave function, \fref{fig:fs_twogauss}(b), by displaying interferences and some population around $x\sim0$ after the wave packets have crossed.
This is clearly a drawback of the classical Schr\"odinger equation.
Let us notice that a very narrow wave-packet (for example, one whose initial support does only include the trajectories with positive initial positions) as discussed in \sref{sec:BMnarrow} will avoid this spurious interference effect of the CSE.

\subsection{Wave packet impinging on a barrier}

Another text-book example of quantum dynamics is a wave packet impinging on a barrier.
In this case, the potential will be modeled by a (Gaussian) barrier of width $\sigma_\textrm{b}$ and height $V_\textrm{b}$ centered at $x_\textrm{b}$, i.e.,
\begin{align} 
V = V_\textrm{b} \exp\left(-\frac{(x-x_\textrm{b})^2}{2 \sigma^2_\textrm{b}}\right) ,
\label{eq:barrier}
\end{align} 
with $\sigma_\textrm{b} = 1$ and $x_\textrm{b} = 0$.
The initial wave function will be described by \eref{eq:gauss} with $x_0 = -4$, $\sigma_x = 1$ and $k_0 = 2.5$, therefore the classical particle will have energy
\begin{align} 
E_\textrm{cl} = \frac{\hbar^2 k_0^2}{2m} = 3.125  .
\end{align}
In the quantum case, however, the total energy contains an additional term,
\begin{align} 
E_\textrm{q} &= -\frac{\hbar^2}{2m} \int_{-\infty}^\infty \Psi^*_G(x) \frac{\partial^2 \Psi_G(x)}{\partial x^2} dx
\nonumber \\ &= \frac{\hbar^2 k_0^2}{2m} + \frac{\hbar^2}{4m\sigma^2} = 3.375,
\end{align} 
and therefore the conditions for transmission/reflection on the barrier will differ slightly.
We will consider two different cases, depending on whether the wave packet has more or less energy than the potential barrier.

\begin{figure}
\centerline{\includegraphics[width=0.99\columnwidth]{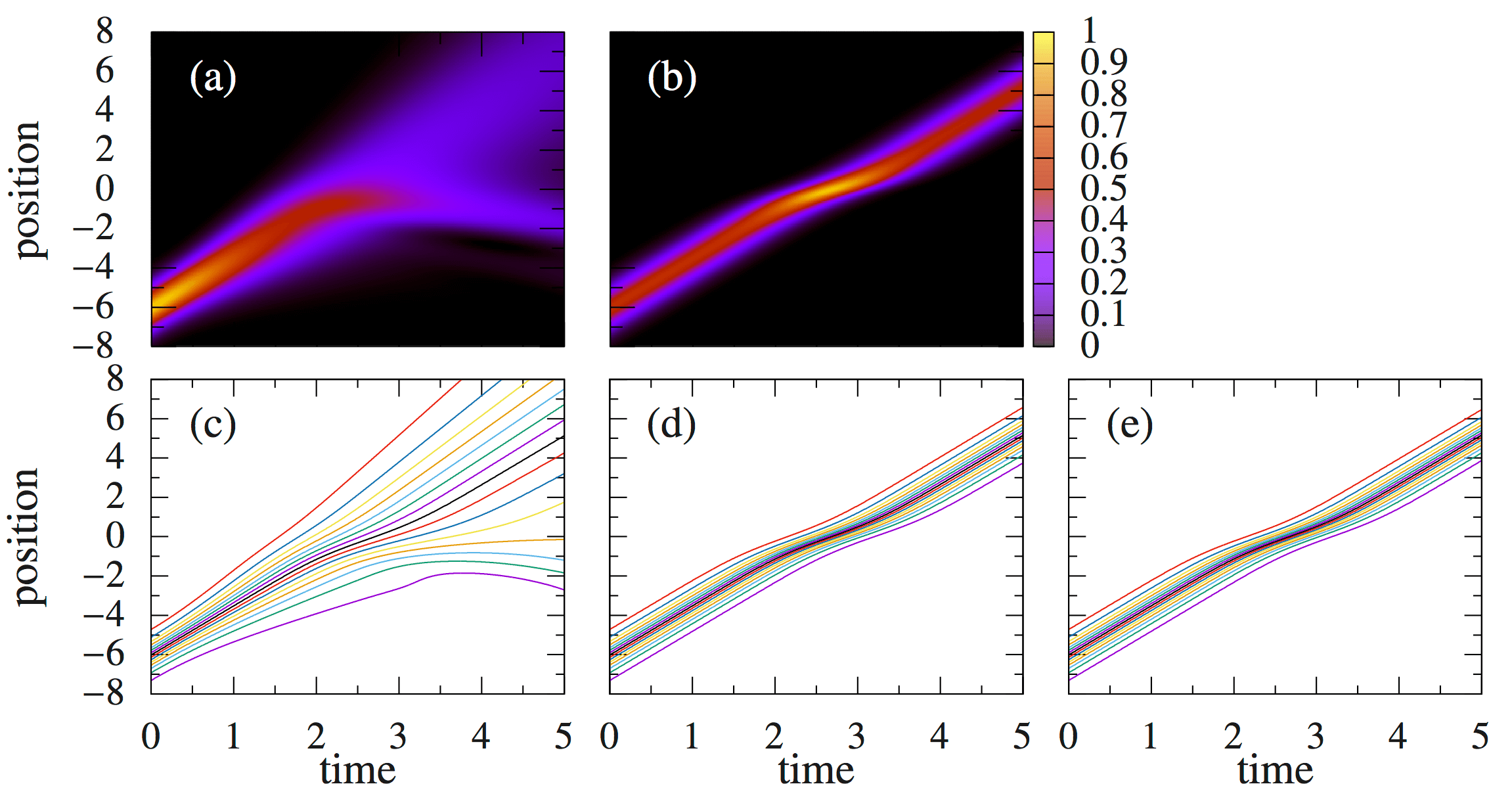}}
\caption{
Same as \fref{fig:fs_gauss} but for a wave packet impinging on a (low) barrier.
The initial wave function is given by \eref{eq:gauss} with $\sigma = 1$, $x_0 = -6$, and $k_0 = 2.5$.
The barrier potential is described by \eref{eq:barrier} with $\sigma_\textrm{b} = 1$, $x_\textrm{b} = 0$, and $V_\textrm{b} = 2$.
\label{fig:barrier_more}
}
\end{figure}

The first of these cases, with $V_\textrm{b} = 2$, is depicted in \fref{fig:barrier_more}.
We know that a quantum particle will get partially reflected and partially transmitted.
This is shown in \fref{fig:barrier_more}(a), where quantum wave function split in two parts with a small fraction getting reflected and the rest going over the barrier.
This is confirmed by the quantum trajectories, \fref{fig:barrier_more}(c), some of which get reflected on the barrier.
Because $E_\textrm{cl} > V_\textrm{b}$, a classical particle will climb up the barrier and slide down the other side without any net change in its energy, as shown by the classical trajectories in \fref{fig:barrier_more}(e).
Again, as these classical trajectories do not cross, the CSE trajectories follow them perfectly, see \fref{fig:barrier_more}(d).
It is interesting to see how this is visualized with the classical wave function, \fref{fig:barrier_more}(b):
the particle slows down and gets narrower around the barrier position, and after transversing it, recovers its previous shape and velocity.

\begin{figure}
\centerline{\includegraphics[width=0.99\columnwidth]{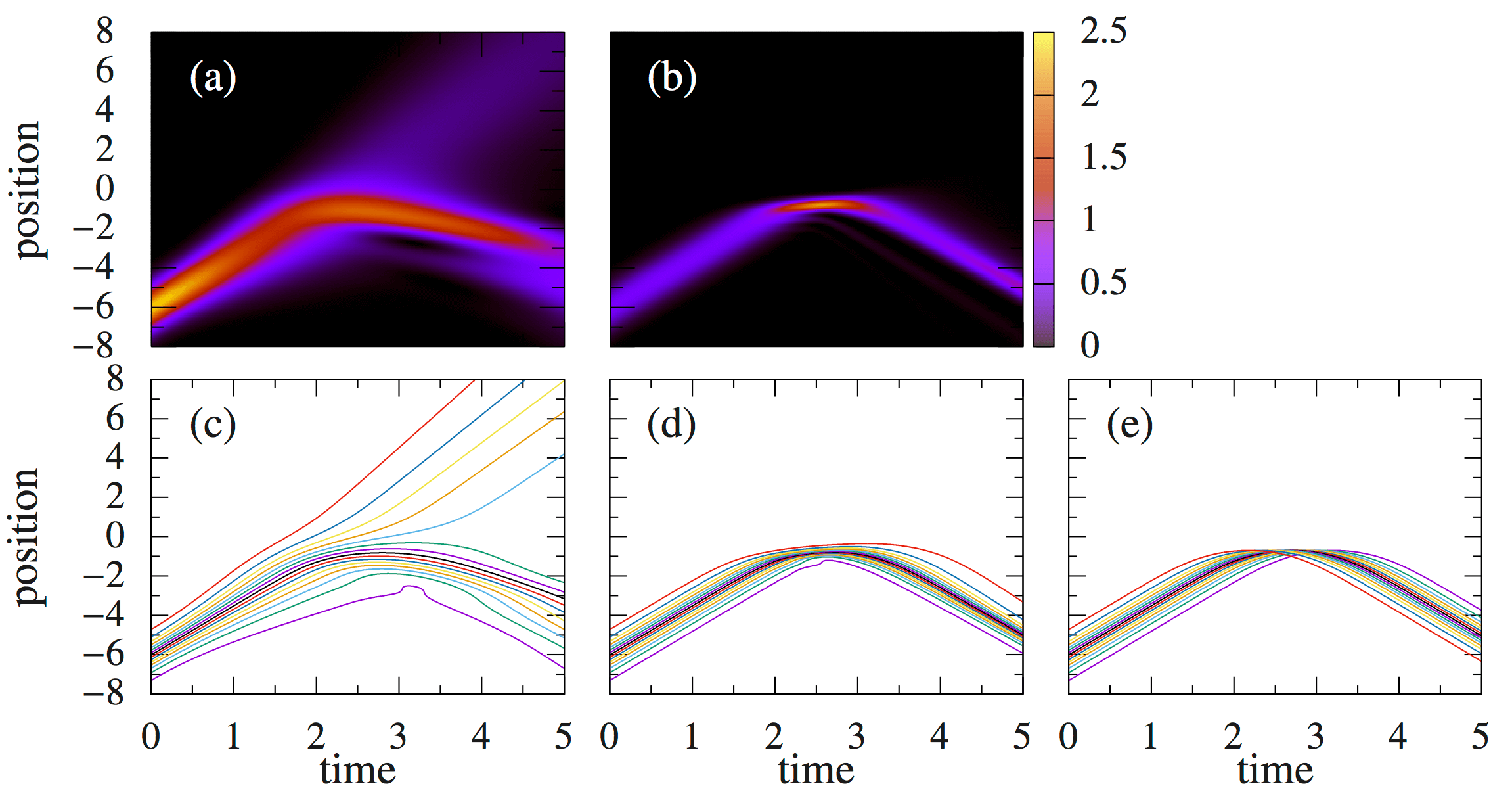}}
\caption{
Same as \fref{fig:fs_gauss} but for a wave packet impinging on a (high) barrier.
The initial wave function is given by \eref{eq:gauss} with $\sigma = 1$, $x_0 = -6$, and $k_0 = 2.5$.
The barrier potential is described by \eref{eq:barrier} with $\sigma_\textrm{b} = 1$, $x_\textrm{b} = 0$, and $V_\textrm{b} = 4$.
\label{fig:barrier_less}
}
\end{figure}

The other case, where $V_\textrm{b} = 4$, is depicted in \fref{fig:barrier_less}.
Again, the quantum particle gets mostly reflected and partially transmitted, seen in both the wave function and the trajectories in \fref{fig:barrier_less}(a,c).
The classical case does not have enough energy to surpass the barrier and it is forced to be completely reflected.
However, one would expect in this case the classical trajectories to cross (the first ones arriving at the barrier should be the first ones reflected), as in \fref{fig:barrier_less}(e).
Therefore, the CSE trajectories equation are forced to deviate from this behavior:
the first ones to arrive at the barrier linger longer at the classical turning point, while the ones which arrive later turn around earlier, see \fref{fig:barrier_less}(d).
This also produces a narrowing of the classical wave packet while it is ``bouncing'' off the barrier, \fref{fig:barrier_less}(b).
Again, narrowing the classical wave packet, as indicated in \sref{sec:BMnarrow}, will make this feature totally insignificant. Finally, let us notice that there is no tunneling in the classical wave function. 

\subsection{Harmonic potential}

\begin{figure}
\centerline{\includegraphics[width=0.99\columnwidth]{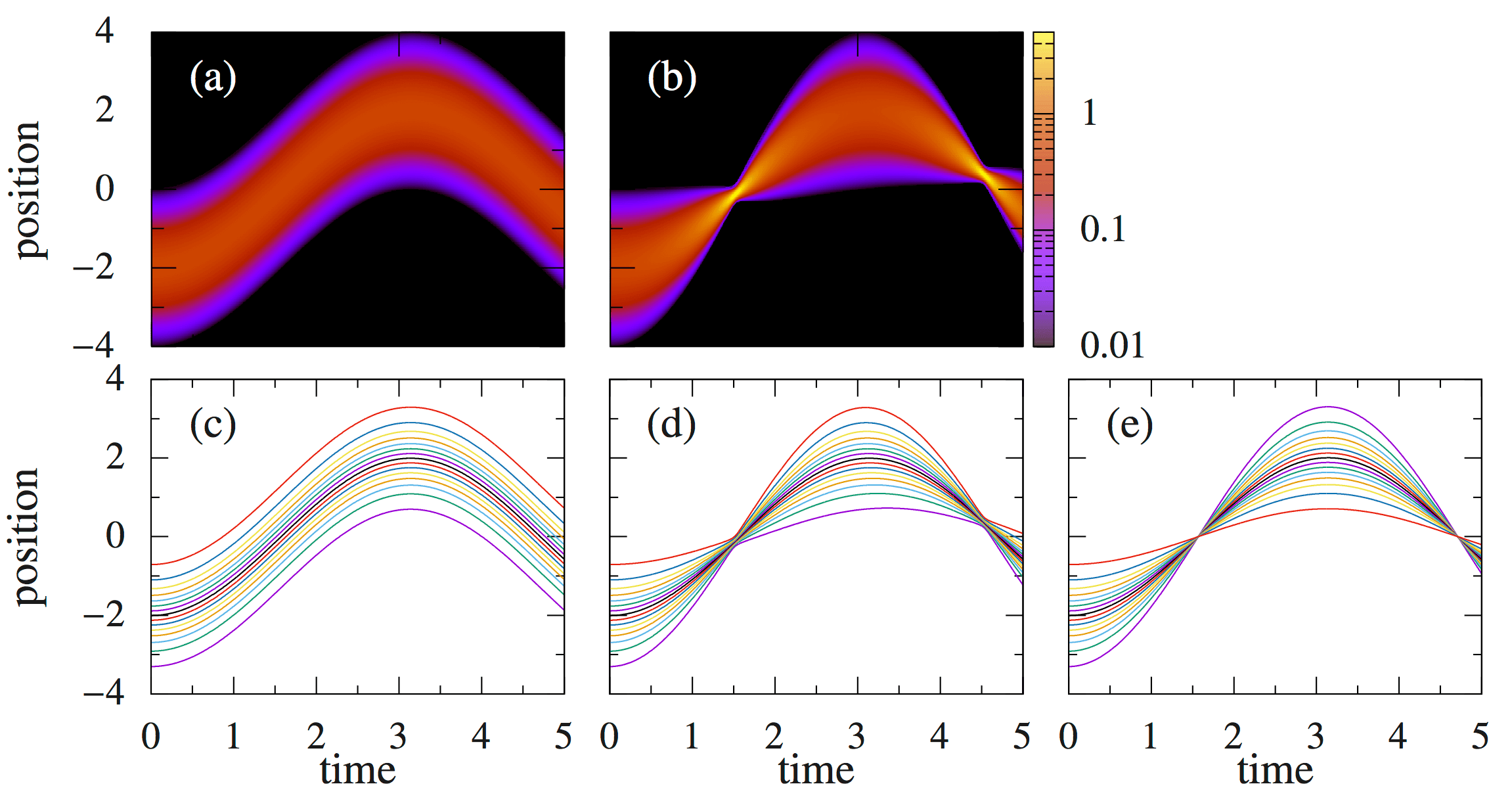}}
\caption{
Same as \fref{fig:fs_gauss} but for a displaced wave packet in a harmonic potential.
The initial wave function is given by \eref{eq:gauss} with $\sigma = 1$, $x_0 = -2$, and $k_0 = 0$.
The harmonic potential is described by \eref{eq:harmonic} with $\omega = 1$.
\label{fig:harm_displaced}
}
\end{figure}

\begin{figure}
\centerline{\includegraphics[width=0.99\columnwidth]{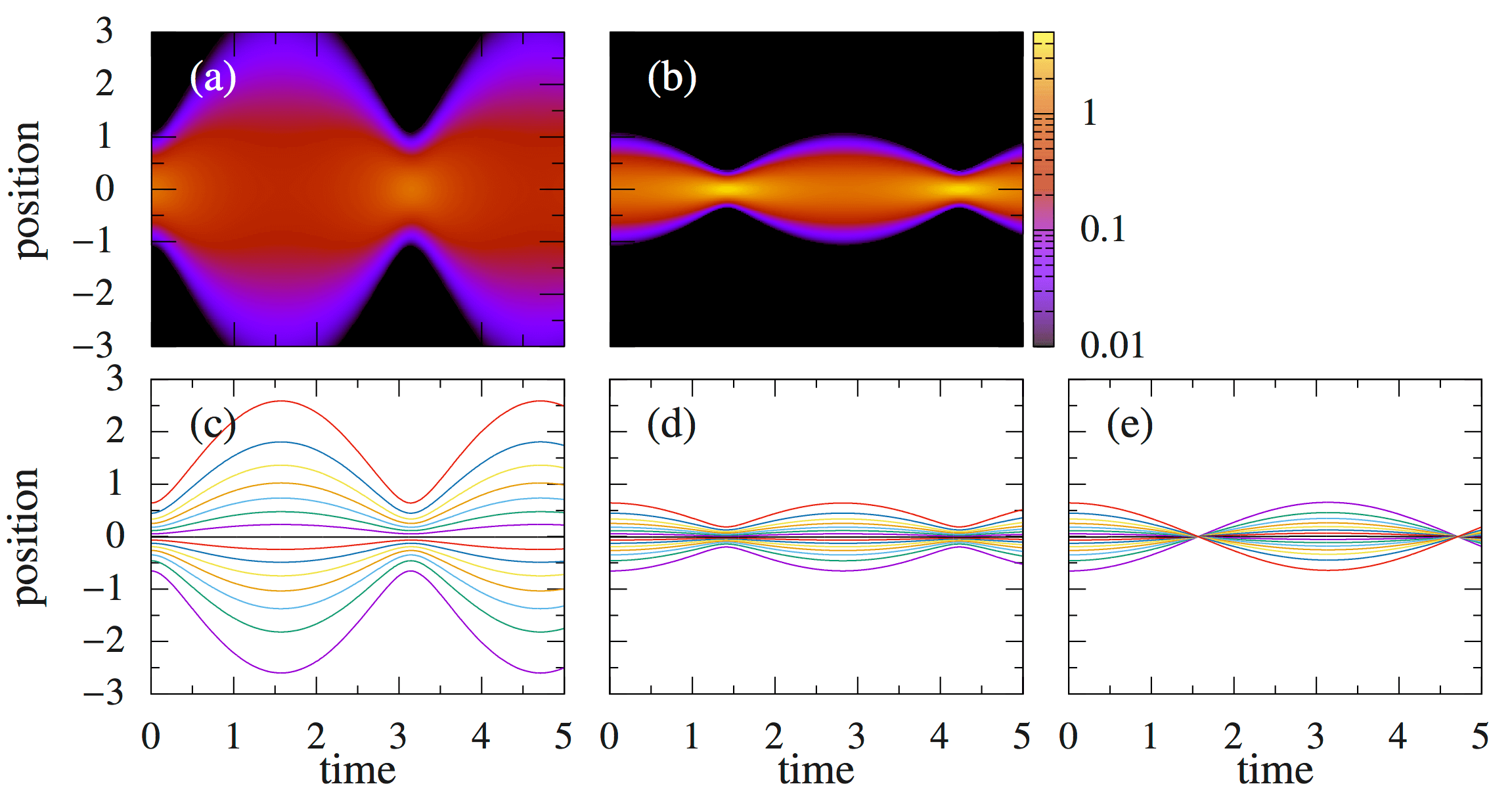}}
\caption{Thinner than ground state.
$\omega=1$
$\sigma=1/2$
$x_0=0$
\label{fig:harm_narrow}
}
\end{figure}

The last case we will study is that of a harmonic oscillator, i.e., a potential of the form
\begin{align} 
V = \frac{1}{2} m \omega x^2 ,
\label{eq:harmonic}
\end{align} 
with an initial Gaussian wave function described by \eref{eq:gauss}.

The first case we study is shown in \fref{fig:harm_displaced}, where the initial wave function is the (quantum) harmonic oscillator ground state (a Gaussian of width 1), but displaced from the minimum of the potential.
The simulation for the (centered) ground state of this system can be found in~\cite{ref3}.
The quantum dynamics show the wave packet not changing shape as it oscillates around the origin, see \fref{fig:harm_displaced}(a).
Therefore, the corresponding trajectories, \fref{fig:harm_displaced}(c), also perform oscillations keeping a fixed distance between them.
On the other hand, the classical trajectories shown in \fref{fig:harm_displaced}(e), all cross the origin at the same time because the oscillation frequency is independent of the oscillation amplitude.
Because of this crossings, the CSE trajectories are again affected by the single-valuedness of $S$, see \fref{fig:harm_displaced}(d):
trajectories starting closer to the center of the trap would oscillate with smaller amplitudes than those starting further away.
In order to avoid the crossing, they seem to bounce off each other around $x \sim 0$.

Similar behavior can be seen in \fref{fig:harm_narrow}, where the initial wave function is now centered around the potential minimum, but it has a width narrower than the harmonic oscillator ground state.
Because the wave packet is narrower than the ground state, the quantum wave packets starts oscillating by expanding and compressing around a width of 1, see \fref{fig:harm_narrow}(a).
The quantum trajectories, \fref{fig:harm_narrow}(c), then follow suit: spreading out when the wave function expands and getting closer together when the wave function compresses.
The classical wave function in \fref{fig:harm_narrow}(b), however, shows the opposite behavior: because classical dynamics should in principle by unaffected by the shape of the wave function, a centered Gaussian will always start getting compressed no matter its initial width.
As in the previous case, we can see the CSE trajectories trying to oscillate around the origin, but, again, because they cannot cross they are forced to repel each other, see \fref{fig:harm_narrow}(d).
This behaviour by the CSE trajectories is similar to the classical trajectories in \fref{fig:harm_narrow}(e).
However, as CSE trajectories cannot cross, they seem to swap places when they are around $x \sim 0$ such that they appear to completely oscillate around the origin (like the classical ones do).
This is also the case for the previous example.
Again, narrowing the classical wave packet, with $N \rightarrow \infty$ which implies $\sigma \rightarrow 0$ for the wave packet of the center of mass as indicated in \eref{variance}, will make this non-classical feature totally insignificant.

\section{Conclusions}
\label{sec:conc}

We have presented a study of the classical Schr\"odinger equation. First, we have shown that Newtonian mechanics leads itself naturally to a reformulation of classical systems in terms of the so-called classical Schr\"odinger equation. We emphasize that it is usually unnoticed that reaching the classical Schr\"odinger equation requires an additional single-valued requirement for the action $S(x,t)$ not present in the original Hamilton-Jacobi formulation. This restriction introduces non-classical (non-Newtonian) features on the trajectories obtained from the classical Schr\"odinger equation. Second, we have shown how the classical Schr\"odinger equation can be reached from  quantum theories by invoking the quantum-to-classical transition. We have shown Rosen's proposal for the recovery of classical formalism from Bohmian mechanics of a single particle quantum system, and then we have presented a generalization of his work applicable to a more general type of experiments where a classical behavior emerges naturally at the center of mass of a quantum object (not at each individual particle) with a large number of identical particles. 

The continuity between quantum and classical systems expressed here is an attempt to bridge the gap between these two theories. By providing a description of quantum mechanics in terms of trajectories, and their encompassing wave functions, Bohmian mechanics provides a common formalism which can be applied to both. The transition from one regime to another under plausible conditions is given in \sref{sec:BM}, where we show that not only the common language makes it \textit{a priori} an adequate vehicle for relating these theories, but that classical dynamics actually emerge naturally from purely quantum behavior. In this work, we warn about two difficulties that have to be acknowledged in this quantum-to-classical Bohmian path. First, the possible unphysical features of the classical Schr\"odinger equation when dealing with ensembles of particles because of imposing a single-valued condition on the action $S(x,t)$. Second, the several arguments that give strong support to the idea that it is better to formulate the quantum-to-classical Bohmian path in terms of the center of mass rather than in terms of individual particles. See the recent work of the authors in Ref.~\cite{ref3}.

\section*{Acknowledgements}

This work has been partially supported by the Fondo Europeo de Desarrollo Regional (FEDER) and Ministerio de Econom\'{i}a y Competitividad through the Spanish Project Nos. TEC2012-31330 and TEC2015-67462-C2-1-R, the Generalitat de Catalunya (2014 SGR-384),  by the European Union Seventh Framework Program under the Grant agreement no: 604391 of the Flagship initiative  ``Graphene-Based Revolutions in ICT and Beyond'' and the Okinawa Institute of Science and Technology Graduate University.


\begin{thebibliography}{99}
\bibitem{footnote1}
The development of the classical formalism done here is for a single-particle in a one-dimensional space. The generalization of the conclusions extracted here to system with higher dimensionality is not trivial. The single-valued $S(x,t)$ in the 1D the classical Schr\"odinger equation, implies that the trajectories do not cross in the physical space. However,  the single-valued Hamilton's principal function  $S(x_1,x_2,x_3,.....,x_N,t)$ in the  $N$-dimensional classical Schr\"odinger equation, implies that the trajectories do not cross in the ($N$-dimensional) configuration space, but such trajectories can indeed cross in physical space. In addition, structures such as vortices (that require physical spaces with dimensions greater than 1) are not present in our 1D analysis.  In any case,  the 1D case is enough to reveal important features and drawbacks of the classical Schr\"odinger equation.
\bibitem{footnote2}
While this particular form is a convenient choice to show the non single-valued property of $S(x,t;\alpha)$, there are of course other possibilities not of this form.
\bibitem{footnote3}
We consider different trajectories belonging to different experiments. All these single-particle experiments are prepared with the same classical wave function $\psi_{cl}(x,t) = R(x,t) \exp(i S(x,t)/\hbar)$.
\bibitem{footnote4}
There are many other arguments to arrive to \eref{bohmvel} without using the analogy with the classical hamilton-Jacobi equation. See, for example~\cite{durr13}.
\bibitem{footnote5}
One could very well argue that $J(x_{cm},t)$ is an additional unknown.
However, one can usually fix $J(x_{cm},t)$ by fixing a conservation of probability (although this is not necessary for a conditional wave function).
Nevertheless, and unlike the single-particle case, the number of unknowns for the many-body system is equal or higher than the number of equations to satisfy.
\bibitem{footnote6}
Here, the \lq\lq{}same\rq\rq{} means that not the same microscopic Bohmian particles, but the same macroscopic appearance of the object.
\end{thebibliography}
\end{document}